\documentclass[11pt,letterpaper]{article}
\usepackage{latexsym}

\makeatletter
                                                                               
\def\icite{\@ifnextchar [{\@tempswatrue\@citey}{\@tempswafalse\@citey[]}}
\def\@citex[#1]#2{%
\if@filesw \immediate \write \@auxout {\string \citation {#2}}\fi
\@tempcntb\m@ne \let\@h@ld\relax \def\@citea{}%
\@cite{%
  \@for \@citeb:=#2\do {%
    \@ifundefined {b@\@citeb}%
      {\@h@ld\@citea\@tempcntb\m@ne{\bf ?}%
      \@warning {Citation `\@citeb ' on page \thepage \space undefined}}%
      {\@tempcnta\@tempcntb \advance\@tempcnta\@ne%
      \@tempcntb\number\csname b@\@citeb \endcsname \relax%
      \ifnum\@tempcnta=\@tempcntb 
        \ifx\@h@ld\relax%
          \edef \@h@ld{\@citea\csname b@\@citeb\endcsname}%
        \else%
          \edef\@h@ld{\ifmmode{-}\else--\fi\csname b@\@citeb\endcsname}%
        \fi%
      \else
        \@h@ld\@citea\csname b@\@citeb \endcsname%
        \let\@h@ld\relax%
      \fi}%
    \def\@citea{,\penalty\@highpenalty\,}%
  }\@h@ld
}{#1}}
\def\@citey[#1]#2{%
\if@filesw \immediate \write \@auxout {\string \citation {#2}}\fi
\@tempcntb\m@ne \let\@h@ld\relax \def\@citea{}%
\@icite{%
  \@for \@citeb:=#2\do {%
    \@ifundefined {b@\@citeb}%
      {\@h@ld\@citea\@tempcntb\m@ne{\bf ?}%
      \@warning {Citation `\@citeb ' on page \thepage \space undefined}}%
      {\@tempcnta\@tempcntb \advance\@tempcnta\@ne%
      \@tempcntb\number\csname b@\@citeb \endcsname \relax%
      \ifnum\@tempcnta=\@tempcntb 
        \ifx\@h@ld\relax%
          \edef \@h@ld{\@citea\csname b@\@citeb\endcsname}%
        \else%
          \edef\@h@ld{\ifmmode{-}\else--\fi\csname b@\@citeb\endcsname}%
        \fi%
      \else
        \@h@ld\@citea\csname b@\@citeb \endcsname%
        \let\@h@ld\relax%
      \fi}%
    \def\@citea{,\penalty\@highpenalty\,}%
  }\@h@ld
}{#1}}
\def\@cite#1#2{{$^{#1}$\if@tempswa , #2\fi }}
\def\@icite#1#2{{$#1$\if@tempswa , #2\fi }}

\gdef\@publabel{\hfil}
\gdef\@pubdate{\null}
\gdef\@pubnumber{\null}
\gdef\@author{\null}
\gdef\@title{\null}
\gdef\@abstract{\null}
\long\def\pubdate#1{\gdef\@pubdate{#1}}
\long\def\pubnumber#1{\gdef\@pubnumber{#1}}
\long\def\publabel#1{\gdef\@publabel{#1}}
\long\def\author#1{\gdef\@author{#1}}
\long\def\title#1{\gdef\@title{#1}}
\long\def\abstract#1{\gdef\@abstract{#1}}
\def\titlerelax{
\let\maketitle\relax
\let\settitleparameters\relax
\let\consolidatetitle\relax
\let\inittitlepage\relax
\let\finishtitlepage\relax
\let\titlepagecontents\relax
\let\multithanks\relax
\let\titlebaselines\relax
\let\@makepub\relax
\let\@maketitle\relax
\let\@makeauthor\relax
\let\@makeabstract\relax
\let\@maketitlenote\relax
\let\thanks\relax
\let\titlerelax\relax}

\def\titleclean
{\gdef\@titlenote{}
\gdef\@abstract{}
\gdef\@author{}
\gdef\@title{}
\gdef\@pubdate{}\gdef\@pubnumber{}\gdef\@publabel{}
\gdef\@dpublabel{}
}
\def\@makepub{\vbox to \z@{\hbox to \textwidth{\hfill
\@publabel \hfill
\llap{\parbox[t]{0.25\textwidth}{\raggedleft\@pubnumber}}}%
\vss}}
\def\@maketitle{\vskip 60pt \begin{center}
 {\LARGE \@title \par}
 \end{center}}
\def\@makeauthor{{%
\def\and{\smallskip {\normalsize \rm and\smallskip }}
\def\And{\medskip {\normalsize \rm and\\}\medskip}
\long\def\address##1{{\def\and{\\and\\}\medskip
				{\small \it \\##1\\}
}}
{\centering
 \vskip 3em
 \large \lineskip .75em
 \@author}
 \par}}
\def\@makedate{\vskip 1.5em
 {\raggedright \small \noindent\@pubdate \par}}
\def\@makeabstract{\vskip 1.5em
{\small
\begin{center}
{\bf ABSTRACT\vspace{-.5em}\vspace{0pt}}
\end{center}
\quotation \@abstract \endquotation}}
\def\maketitle{\titlepage
\let\footnotesize\small \setcounter{page}{0}
\def\thefootnote{\arabic{footnote}}
\@makepub
\vfil
\@maketitle
\@makeauthor
\vfil
\@makeabstract
\@thanks
\vfil
\@makedate
\if@restonecol\twocolumn \else \eject \fi
\titlerelax \titleclean
\def\thefootnote{\alph{footnote}}
\setcounter{footnote}{0}
}

\ifcase\@ptsize
 \font\tenmsa=msam10
 \font\sevenmsa=msam7
 \font\fivemsa=msam5
 \font\tenmsb=msbm10
 \font\sevenmsb=msbm7
 \font\fivemsb=msbm5
\or
 \font\tenmsa=msam10 scaled \magstephalf
 \font\sevenmsa=msam8
 \font\fivemsa=msam6
 \font\tenmsb=msbm10 scaled \magstephalf
 \font\sevenmsb=msbm8
 \font\fivemsb=msbm6
\or
 \font\tenmsa=msam10 scaled \magstep1
 \font\sevenmsa=msam8
 \font\fivemsa=msam6
 \font\tenmsb=msbm10 scaled \magstep1
 \font\sevenmsb=msbm8
 \font\fivemsb=msbm6
\fi

\newfam\msafam
\newfam\msbfam
\textfont\msafam=\tenmsa  \scriptfont\msafam=\sevenmsa
  \scriptscriptfont\msafam=\fivemsa
\textfont\msbfam=\tenmsb  \scriptfont\msbfam=\sevenmsb
  \scriptscriptfont\msbfam=\fivemsb

\def\hexnumber@#1{\ifnum#1<10 \number#1\else
 \ifnum#1=10 A\else\ifnum#1=11 B\else\ifnum#1=12 C\else
 \ifnum#1=13 D\else\ifnum#1=14 E\else\ifnum#1=15 F\fi\fi\fi\fi\fi\fi\fi}

\def\msa@{\hexnumber@\msafam}
\def\msb@{\hexnumber@\msbfam}
\mathchardef\boxdot="2\msa@00
\mathchardef\boxplus="2\msa@01
\mathchardef\boxtimes="2\msa@02
\mathchardef\square="0\msa@03
\mathchardef\blacksquare="0\msa@04
\mathchardef\centerdot="2\msa@05
\mathchardef\lozenge="0\msa@06
\mathchardef\blacklozenge="0\msa@07
\mathchardef\circlearrowright="3\msa@08
\mathchardef\circlearrowleft="3\msa@09
\mathchardef\rightleftharpoons="3\msa@0A
\mathchardef\leftrightharpoons="3\msa@0B
\mathchardef\boxminus="2\msa@0C
\mathchardef\Vdash="3\msa@0D
\mathchardef\Vvdash="3\msa@0E
\mathchardef\vDash="3\msa@0F
\mathchardef\twoheadrightarrow="3\msa@10
\mathchardef\twoheadleftarrow="3\msa@11
\mathchardef\leftleftarrows="3\msa@12
\mathchardef\rightrightarrows="3\msa@13
\mathchardef\upuparrows="3\msa@14
\mathchardef\downdownarrows="3\msa@15
\mathchardef\upharpoonright="3\msa@16

\mathchardef\downharpoonright="3\msa@17
\mathchardef\upharpoonleft="3\msa@18
\mathchardef\downharpoonleft="3\msa@19
\mathchardef\rightarrowtail="3\msa@1A
\mathchardef\leftarrowtail="3\msa@1B
\mathchardef\leftrightarrows="3\msa@1C
\mathchardef\rightleftarrows="3\msa@1D
\mathchardef\Lsh="3\msa@1E
\mathchardef\Rsh="3\msa@1F
\mathchardef\rightsquigarrow="3\msa@20
\mathchardef\leftrightsquigarrow="3\msa@21
\mathchardef\looparrowleft="3\msa@22
\mathchardef\looparrowright="3\msa@23
\mathchardef\circeq="3\msa@24
\mathchardef\succsim="3\msa@25
\mathchardef\gtrsim="3\msa@26
\mathchardef\gtrapprox="3\msa@27
\mathchardef\multimap="3\msa@28
\mathchardef\therefore="3\msa@29
\mathchardef\because="3\msa@2A
\mathchardef\doteqdot="3\msa@2B

\mathchardef\triangleq="3\msa@2C
\mathchardef\precsim="3\msa@2D
\mathchardef\lesssim="3\msa@2E
\mathchardef\lessapprox="3\msa@2F
\mathchardef\eqslantless="3\msa@30
\mathchardef\eqslantgtr="3\msa@31
\mathchardef\curlyeqprec="3\msa@32
\mathchardef\curlyeqsucc="3\msa@33
\mathchardef\preccurlyeq="3\msa@34
\mathchardef\leqq="3\msa@35
\mathchardef\leqslant="3\msa@36
\mathchardef\lessgtr="3\msa@37
\mathchardef\backprime="0\msa@38
\mathchardef\risingdotseq="3\msa@3A
\mathchardef\fallingdotseq="3\msa@3B
\mathchardef\succcurlyeq="3\msa@3C
\mathchardef\geqq="3\msa@3D
\mathchardef\geqslant="3\msa@3E
\mathchardef\gtrless="3\msa@3F
\mathchardef\sqsubset="3\msa@40
\mathchardef\sqsupset="3\msa@41
\mathchardef\vartriangleright="3\msa@42
\mathchardef\vartriangleleft="3\msa@43
\mathchardef\trianglerighteq="3\msa@44
\mathchardef\trianglelefteq="3\msa@45
\mathchardef\bigstar="0\msa@46
\mathchardef\between="3\msa@47
\mathchardef\blacktriangledown="0\msa@48
\mathchardef\blacktriangleright="3\msa@49
\mathchardef\blacktriangleleft="3\msa@4A
\mathchardef\vartriangle="3\msa@4D
\mathchardef\blacktriangle="0\msa@4E
\mathchardef\triangledown="0\msa@4F
\mathchardef\eqcirc="3\msa@50
\mathchardef\lesseqgtr="3\msa@51
\mathchardef\gtreqless="3\msa@52
\mathchardef\lesseqqgtr="3\msa@53
\mathchardef\gtreqqless="3\msa@54
\mathchardef\Rrightarrow="3\msa@56
\mathchardef\Lleftarrow="3\msa@57
\mathchardef\veebar="2\msa@59
\mathchardef\barwedge="2\msa@5A
\mathchardef\doublebarwedge="2\msa@5B
\mathchardef\angle="0\msa@5C
\mathchardef\measuredangle="0\msa@5D
\mathchardef\sphericalangle="0\msa@5E
\mathchardef\varpropto="3\msa@5F
\mathchardef\smallsmile="3\msa@60
\mathchardef\smallfrown="3\msa@61
\mathchardef\Subset="3\msa@62
\mathchardef\Supset="3\msa@63
\mathchardef\Cup="2\msa@64

\mathchardef\Cap="2\msa@65

\mathchardef\curlywedge="2\msa@66
\mathchardef\curlyvee="2\msa@67
\mathchardef\leftthreetimes="2\msa@68
\mathchardef\rightthreetimes="2\msa@69
\mathchardef\subseteqq="3\msa@6A
\mathchardef\supseteqq="3\msa@6B
\mathchardef\bumpeq="3\msa@6C
\mathchardef\Bumpeq="3\msa@6D
\mathchardef\lll="3\msa@6E

\mathchardef\ggg="3\msa@6F

\mathchardef\circledS="0\msa@73
\mathchardef\pitchfork="3\msa@74
\mathchardef\dotplus="2\msa@75
\mathchardef\backsim="3\msa@76
\mathchardef\backsimeq="3\msa@77
\mathchardef\complement="0\msa@7B
\mathchardef\intercal="2\msa@7C
\mathchardef\circledcirc="2\msa@7D
\mathchardef\circledast="2\msa@7E
\mathchardef\circleddash="2\msa@7F
\def\ulcorner{\delimiter"4\msa@70\msa@70 }
\def\urcorner{\delimiter"5\msa@71\msa@71 }
\def\llcorner{\delimiter"4\msa@78\msa@78 }
\def\lrcorner{\delimiter"5\msa@79\msa@79 }
\def\yen{\mathhexbox\msa@55 }
\def\checkmark{\mathhexbox\msa@58 }
\def\circledR{\mathhexbox\msa@72 }
\def\maltese{\mathhexbox\msa@7A }
\mathchardef\lvertneqq="3\msb@00
\mathchardef\gvertneqq="3\msb@01
\mathchardef\nleq="3\msb@02
\mathchardef\ngeq="3\msb@03
\mathchardef\nless="3\msb@04
\mathchardef\ngtr="3\msb@05
\mathchardef\nprec="3\msb@06
\mathchardef\nsucc="3\msb@07
\mathchardef\lneqq="3\msb@08
\mathchardef\gneqq="3\msb@09
\mathchardef\nleqslant="3\msb@0A
\mathchardef\ngeqslant="3\msb@0B
\mathchardef\lneq="3\msb@0C
\mathchardef\gneq="3\msb@0D
\mathchardef\npreceq="3\msb@0E
\mathchardef\nsucceq="3\msb@0F
\mathchardef\precnsim="3\msb@10
\mathchardef\succnsim="3\msb@11
\mathchardef\lnsim="3\msb@12
\mathchardef\gnsim="3\msb@13
\mathchardef\nleqq="3\msb@14
\mathchardef\ngeqq="3\msb@15
\mathchardef\precneqq="3\msb@16
\mathchardef\succneqq="3\msb@17
\mathchardef\precnapprox="3\msb@18
\mathchardef\succnapprox="3\msb@19
\mathchardef\lnapprox="3\msb@1A
\mathchardef\gnapprox="3\msb@1B
\mathchardef\nsim="3\msb@1C
\mathchardef\napprox="3\msb@1D
\mathchardef\varsubsetneq="3\msb@20
\mathchardef\varsupsetneq="3\msb@21
\mathchardef\nsubseteqq="3\msb@22
\mathchardef\nsupseteqq="3\msb@23
\mathchardef\subsetneqq="3\msb@24
\mathchardef\supsetneqq="3\msb@25
\mathchardef\varsubsetneqq="3\msb@26
\mathchardef\varsupsetneqq="3\msb@27
\mathchardef\subsetneq="3\msb@28
\mathchardef\supsetneq="3\msb@29
\mathchardef\nsubseteq="3\msb@2A
\mathchardef\nsupseteq="3\msb@2B
\mathchardef\nparallel="3\msb@2C
\mathchardef\nmid="3\msb@2D
\mathchardef\nshortmid="3\msb@2E
\mathchardef\nshortparallel="3\msb@2F
\mathchardef\nvdash="3\msb@30
\mathchardef\nVdash="3\msb@31
\mathchardef\nvDash="3\msb@32
\mathchardef\nVDash="3\msb@33
\mathchardef\ntrianglerighteq="3\msb@34
\mathchardef\ntrianglelefteq="3\msb@35
\mathchardef\ntriangleleft="3\msb@36
\mathchardef\ntriangleright="3\msb@37
\mathchardef\nleftarrow="3\msb@38
\mathchardef\nrightarrow="3\msb@39
\mathchardef\nLeftarrow="3\msb@3A
\mathchardef\nRightarrow="3\msb@3B
\mathchardef\nLeftrightarrow="3\msb@3C
\mathchardef\nleftrightarrow="3\msb@3D
\mathchardef\divideontimes="2\msb@3E
\mathchardef\varnothing="0\msb@3F
\mathchardef\nexists="0\msb@40
\mathchardef\mho="0\msb@66
\mathchardef\thorn="0\msb@67
\mathchardef\beth="0\msb@69
\mathchardef\gimel="0\msb@6A
\mathchardef\daleth="0\msb@6B
\mathchardef\lessdot="3\msb@6C
\mathchardef\gtrdot="3\msb@6D
\mathchardef\ltimes="2\msb@6E
\mathchardef\rtimes="2\msb@6F
\mathchardef\shortmid="3\msb@70
\mathchardef\shortparallel="3\msb@71
\mathchardef\smallsetminus="2\msb@72
\mathchardef\thicksim="3\msb@73
\mathchardef\thickapprox="3\msb@74
\mathchardef\approxeq="3\msb@75
\mathchardef\succapprox="3\msb@76
\mathchardef\precapprox="3\msb@77
\mathchardef\curvearrowleft="3\msb@78
\mathchardef\curvearrowright="3\msb@79
\mathchardef\digamma="0\msb@7A
\mathchardef\varkappa="0\msb@7B
\mathchardef\hslash="0\msb@7D
\mathchardef\hbar="0\msb@7E
\mathchardef\backepsilon="3\msb@7F
\def\Bbb{\ifmmode\let\next\Bbb@\else
 \def\next{\errmessage{Use \string\Bbb\space only in math mode}}\fi\next}
\def\Bbb@#1{{\Bbb@@{#1}}}
\def\Bbb@@#1{\fam\msbfam#1}

\def\bk {{\hskip 0.2 cm}}

\def\acknowledgements{\@startsection{section}{4}
{\z@}{-3.5ex plus -1ex minus -.2ex}{2.3ex plus .2ex}{\normalsize\bf}
{Acknowledgements}}

\newcommand{\tab}[1]{{\sc Tab.}\,{\sf #1}}	  
\newcommand{\eq}[1]{{\sc Eq.}\,{\sf (#1)}}	  
\newcommand{\refoth}[1]{{\sf #1}}  	  	  

\def\bbbz {\Bbb{Z}}		  

\def\bbbn {\Bbb{N}}   	          
\def\bbbc {\Bbb{C}}

\newtheorem{definition}{Definition}[section]
\newtheorem{theorem}[definition]{Theorem}
\newtheorem{lemma}[definition]{Lemma}
\newtheorem{proposition}[definition]{Proposition}
\newcounter{defs}[section]

\newcommand{\be}{\begin{equation}}
\newcommand{\ee}{\end{equation}}
\newcommand{\bea}{\begin{eqnarray}}
\newcommand{\eea}{\end{eqnarray}}

\newcommand{\bdf}{\stepcounter{defs}\begin{definition}}
\newcommand{\edf}{\end{definition}}
\newcommand{\bth}{\stepcounter{defs}\begin{theorem}}
\newcommand{\eth}{\end{theorem}}
\newcommand{\blm}{\stepcounter{defs}\begin{lemma}}
\newcommand{\elm}{\end{lemma}}
\newcommand{\bpr}{\stepcounter{defs}\begin{proposition}}
\newcommand{\epr}{\end{proposition}}
\newcommand{\bprf}{Proof: }
\newcommand{\eprf}{\hfill $\Box$ \\}
\newcounter{pics}
\newcommand{\bpic}[4]{\begin{center}\begin{picture}(#1,#2)(#3,#4)
\refstepcounter{pics}}
\renewcommand{\thepics}{{\sf\roman{pics}}}
\newcommand{\epic}[1]{\end{picture}\\
{\small {\sc Fig.} \thepics \bk #1} \end{center}}
\newcommand{\epicspl}{\end{picture}\\		
\addtocounter{pics}{-1}\end{center}}		

\renewcommand{\thefootnote}{\rm{\alph{footnote}}}
\newcounter{tabs}
\newcommand{\btab}[1]{\refstepcounter{tabs}\begin{center}
\begin{tabular}{#1}}
\renewcommand{\thetabs}{{\sf\alph{tabs}}}
\newcommand{\etab}[1]{\end{tabular}\\[1.5ex]
{\small {\sc Tab.} \thetabs \bk #1} \end{center}}

\def\noi {\noindent}

\newcommand{\ket}[1]{\left| {#1} \right\rangle}	
\newcommand{\spn}[1]{{\rm span}\{{#1}\}}	
\newcommand{\vm}[1]{{\langle #1 \rangle}}	

\def\pmb#1{\setbox0=\hbox{#1}%
 \kern-.025em\copy0\kern-\wd0
 \kern.05em\copy0\kern-\wd0
 \kern-.025em\raise.0433em\box0 }

\def\cQ{{\cal Q}}
\def\cG{{\cal G}}
\def\cL{{\cal L}}
\def\cH{{\cal H}}

\makeatother

\def\ta{{\sf T}}       
\def\vm{{\cal V}}        
\def\vir{{\sf V}}        
\def\salg{{\cal A}}      
\def\ordering{{\cal O}}  
\def\cset{{\cal C}}      
\newcommand{\osm}[1]{{{<}_{{}_{#1}}}}           

\title{The Adapted Ordering Method in Representation Theory}
\author{Beatriz Gato-Rivera\thanks{bgator@imaff.cfmac.csic.es}
\address{Instituto de Matem\'aticas y F\'\i sica Fundamental, CSIC,\\
Serrano 123, Madrid 28006, Spain \\[.3cm]
NIKHEF-H, Kruislaan 409, NL-1098 SJ Amsterdam, The Netherlands}}
\pubdate{December 2004}
\pubnumber{IMAFF-FM-04/19\\ NIKHEF-2004-012\\ hep-th/0410172}

\abstract{In 1998 the Adapted Ordering Method was developed for the
representation theory of the superconformal algebras. This method, which
proves to be very powerful, can be applied to most algebras and superalgebras,
however. It allows: to determine maximal dimensions for a given type of
singular vector space, to identify all singular vectors by only a few coefficients,
to spot subsingular vectors and to set the basis for constructing embedding
diagrams. In this article we present the
Adapted Ordering Method for general algebras and superalgebras 
which admit a triangulation and review briefly the results obtained for the 
Virasoro algebra and for the $N=2$ and Ramond $N=1$ superconformal algebras.}

\begin{document}

\maketitle



\section{Introduction}

In 1998 the Adapted Ordering Method was developed, by M. D\"{o}rrzapf
and B. Gato-Rivera, for the study of the representation theory of the 
superconformal algebras\cite{SD1}. This method was applied successfully to 
the $N=2$ superconformal algebras\cite{SD1,SD2} (topological, 
Neveu-Schwarz, Ramond and twisted) and to the Ramond $N=1$ superconformal 
algebra\cite{Ramond},
allowing to obtain rigorous proofs for several conjectured results, 
as well as many new results, especially for the case of the twisted $N=2$ 
superconformal algebra and the case of the Ramond $N=1$
superconformal algebra. The Adapted Ordering Method, which
proves to be very powerful, can be applied to most algebras and superalgebras, 
however. It allows: to determine maximal dimensions for a given type of
singular vector space, to identify all singular vectors by only a few coefficients,
to spot subsingular vectors and to set the basis for constructing embedding
diagrams. The purpose of this article is precisely to fill the existing gap, 
providing the description of the Adapted Ordering Method for a general 
algebra or superalgebra which admits a triangular decomposition into 
creation, annihilation and zero mode operators, like is the case for most
Lie algebras and superalgebras\cite{Moody}.

For the given algebra or superalgebra one defines freely generated modules 
over a highest weight vector, denoted as {\it Verma modules}. A Verma 
module is in general not irreducible, but it contains submodules which are
freely generated over, at least, one highest weight vector different from the 
highest weight vector of the Verma module. These vectors are usually 
referred to as {\it singular vectors}. The irreducible highest weight
representations are then obtained as the quotients of the Verma modules
divided by all their submodules. Surprisingly, the complete set of singular
vectors do not generate all the submodules in the case of Verma modules 
which contain {\it subsingular vectors}. The reason is that subsingular 
vectors are singular vectors of the quotient space, but not of the Verma 
module itself\cite{subsing,beatriz1,DGR1,beatriz2}. In this case 
one has to divide further by the submodules generated by the subsingular 
vectors, repeating this division procedure successively, if necessary. 

On the Verma modules one introduces a hermitian contravariant form. 
The vanishing of the corresponding determinant indicates the existence of 
at least one singular vector. The determinant may not detect the whole set
of singular vectors, however, neither does it give the dimension of the space 
of singular vectors with some given weights. There could be in fact more than
one linearly independent singular vectors with the same weights. Therefore,
the dimensions of the spaces of singular vectors have to be found by an
independent procedure, like the Adapted Ordering Method which puts 
upper limits on the dimensions of these singular spaces. For most 
weight spaces of a Verma module these upper limits on the dimensions are
trivial and, as a consequence, we obtain a rigorous proof that there cannot 
exist any singular vectors for these weights. For some weights, however, one 
may find necessary conditions that allow singular vector spaces to exist, either 
only one-dimensional, as is the case for the Virasoro algebra, or even higher 
dimensional spaces, as it happens for the $N=2$ and Ramond $N=1$
superconformal algebras\cite{SD2,Ramond,beatriz2,thesis,paper2}. 

The idea for developing  the Adapted Ordering Method originated, 
in rudimentary form, from
a procedure due to A. Kent for the study of the representations of the
Virasoro algebra\cite{adrian1}. For this purpose the author analytically
continued the Virasoro Verma modules, yielding `generalised' Verma
modules, where he constructed generalised singular vectors expressions
in terms of analytically continued Virasoro operators. This analytical 
continuation is not necessary, however, for the Adapted Ordering Method,
nor is it necessary to construct singular vectors in order to apply it. The
underlying idea is the concept of {\it adapted orderings} for all the possible
terms of the `would be' singular vectors. An adapted ordering is a criterion,
satisfying certain requirements, to decide which of two given terms is the
bigger one. 

In what follows, in section 2 we will describe the Adapted Ordering Method for 
a general algebra or superalgebra which admits a triangulation and, as an 
example, we will apply this method to the Virasoro algebra. In section 3 we will 
review briefly the results obtained for the $N=2$ and the Ramond $N=1$ 
superconformal algebras, as an illustration of the 
power of this method. Section 4 is devoted to conclusions.


\section{The Adapted Ordering Method}
\label{sec:AOM}

Let $\salg$ denote an algebra or superalgebra which admits a triangular 
decomposition into creation, annihilation and zero mode operators:
$\salg = \salg^- \oplus \salg^+ \oplus \salg^0 $, and let $U(\salg)$ be the 
universal enveloping algebra of $\salg$, that also decomposes as
$U(\salg)= U(\salg)^- \oplus U(\salg)^+ \oplus U(\salg)^0 $. The Cartan 
subalgebra ${\cal H}_{\salg}$ is contained in $\salg^0$ but does not
need to be identical to $\salg^0$. In general, an eigenvector with respect
to the Cartan subalgebra with weights given by the set $\{l_i\}$, in 
particular a singular vector $\Psi_{\{l_i\}}$, can be expressed as a sum
of products of creation operators acting on a highest weight vector with
weights $\{\Delta_i\} $: 
\bea
\Psi_{\{l_i\}}&=& \sum_{m_1,m_2,....\in\bbbn_0}^{ }
\sum_{a,b,c,...} k_{a_{-1}^{m_1},a_{-2}^{m_2},...b_{-1}^{n_1},b_{-2}^{n_2},.....} \, 
X_ {\{l_i\}}^{a_{-1}^{m_1},a_{-2}^{m_2},...b_{-1}^{n_1},b_{-2}^{n_2},.....}
\ket{\{\Delta_i\}} \,, \label{eq:psil}
\eea
where $a_{-1}, a_{-2},..... b_{-1}, b_{-2},.....$ are the creation operators of the
algebra (some zero mode operators should also be included if they act as
creation operators),  
$X_ {\{l_i\}}^{a_{-1}^{m_1},a_{-2}^{m_2},...b_{-1}^{n_1},b_{-2}^{n_2},.....} $ are 
the products of the creation operators: $a_{-1}^{m_1}  a_{-2}^{m_2} 
..... b_{-1}^{n_1}  b_{-2}^{n_2}.....$,
with total weights $\{l_i\}$, which will be denoted simply as {\it terms},
and $k_{a_{-1}^{m_1},a_{-2}^{m_2},...b_{-1}^{n_1},b_{-2}^{n_2},.....} \in\bbbc$ 
are coefficients which depend on the given term. A non-trivial term Y then refers
to a term with non-trivial coefficient $k_Y$.

Now let us define the set $\cset_{\{l_i\}}$  as the set of all the terms with 
weights $\{l_i\}$:
\bea
\cset_{\{l_i\}} &=
& \{X_ {\{l_i\}}^{a_{-1}^{m_1},a_{-2}^{m_2},...b_{-1}^{n_1},b_{-2}^{n_2},.....} , 
\, m_1, m_2,.... n_1, n_2,.....\in\bbbn_0 \} \,,
\eea
and let $\ordering$ denote a total ordering on $\cset_{\{l_i\}}$ with global 
minimum, that is an ordering such that any two different terms in 
$\cset_{\{l_i\}}$ are ordered with respect to each other.
Thus $\Psi_{\{l_i\}}$ in \eq{\ref{eq:psil}} needs to contain an
$\ordering$-{\it smallest} $X_0\in\cset_{\{l_i\}}$ with $k_{X_0}\neq 0$ and 
$k_Y=0$ for all $Y\in\cset_{\{l_i\}}$ with $Y\osm{\ordering} X_0$ and 
$Y\neq X_0$. We define an {\it adapted ordering} on $\cset_{\{l_i\}}$ as
follows:

\bdf \label{def:adapt}
A total ordering $\ordering$ on $\cset_{\{l_i\}}$ with global minimum is called 
adapted to the subset $\cset^{A}_{\{l_i\}}\subset\cset_{\{l_i\}}$ in the Verma 
module $\vm_{\{\Delta_i \}}$ if for any element $X_0\in\cset^{A}_{\{l_i\}}$ at 
least one annihilation operator $\Gamma $ exists for which
$\Gamma \, X_0 \ket{\{\Delta_i \}}$ contains a non-trivial term $\tilde{X}$
\bea
\Gamma \, X_0 \ket{\{\Delta_i \}} &=&
( k_{\tilde{X}} \tilde{X} + ....... ) \, \ket{\{\Delta_i \}}
\label{eq:adapt1}
\eea
which is absent, however, for all $\Gamma \, X \ket{\{\Delta_i \}}$,
where $X$ is any term $X \in\cset_{\{l_i\}}$ such that
$X_0 \osm{\ordering} X$. The complement of $\cset^{A}_{\{l_i\}}$,
${\ }\cset^{K}_{\{l_i\}}=\cset_{\{l_i\}} \setminus \cset^{A}_{\{l_i\}}$ 
is the kernel with respect to the ordering $\ordering$ in the Verma 
module $\vm_{\{\Delta_i \}}$.
\edf
In this definition $\Gamma$ should also include zero mode operators 
if they act as annihilation operators.
A crucial point now is that one needs to find suitable, clever orderings in
order to obtain the smallest possible kernels. The reason is that the size
of the kernel puts a limit on the dimension of the corresponding singular
vector space, as stated in the following theorem:

\bth \label{th:dims}
Let $\ordering$ denote an adapted ordering on $\cset^{A}_{\{l_i\}}$ at 
weights $\{l_i\}$ with kernel $\cset^{K}_{\{l_i\}}$ for a given Verma module 
$\vm_{\{\Delta_i \}}$. If the ordering kernel $\cset^{K}_{\{l_i\}}$ has $n$ 
elements, then there are at most $n$ linearly independent singular vectors
$\Psi_{\{l_i\}}$ in $\vm_{\{\Delta_i \}}$ with weights $\{l_i\}$.
\eth

Observe that the maximal possible dimension $n$ does not imply that
all the singular vectors of the corresponding type are $n$-dimensional.
From this theorem one deduces that if $\cset^{K}_{\{l_i\}}=\emptyset$ for
a given Verma module, then there are no singular vectors with weights
$\{l_i\}$ in it. That is:

\bth \label{th:striv}
Let $\ordering$ denote an adapted ordering on $\cset^{A}_{\{l_i\}}$ at 
weights $\{l_i\}$ with trivial kernel $\cset^{K}_{\{l_i\}}=\emptyset$ for a given 
Verma module $\vm_{\{\Delta_i \}}$. A singular vector $\Psi_{\{l_i\}}$ 
in $\vm_{\{\Delta_i \}}$ with weights $\{l_i\}$ is therefore trivial. 
\eth

In addition, the coefficients with respect to the ordering kernel 
$\cset^{K}_{\{l_i\}}$ uniquely identify a singular vector. Since the size
of the ordering kernels are small (one or two, rarely three, terms),
it turns out that a few (one, two, ....) coefficients completely determine
a singular vector no matter its size, what allows to find easily product
expressions for descendant singular vectors. This is summarized in the 
following theorem:

\bth \label{th:kernel}
Let $\ordering$ denote an adapted ordering on $\cset^{A}_{\{l_i\}}$ at 
weights $\{l_i\}$ with kernel $\cset^{K}_{\{l_i\}}$ for a given Verma module 
$\vm_{\{\Delta_i \}}$. If two singular vectors $\Psi^{1}_{\{l_i\}}$ and 
$\Psi^{2}_{\{l_i\}}$ with the same weights $\{l_i\}$ have $k_{X}^1=k_{X}^2$ 
for all $X\in\cset^{K}_{\{l_i\}}$, then
\bea
\Psi^{1}_{\{l_i\}} &\equiv & \Psi^{2}_{\{l_i\}}\,.
\eea
\eth

\bprf
Let us consider ${\Psi}_{\{l_i\}} = \Psi^{1}_{\{l_i\}} - \Psi^{2}_{\{l_i\}}$, which
does not contain any terms of the ordering kernel $\cset^K_{\{l_i\}}$, simply 
because $k_{X}^1=k_{X}^2$ for all $X\in\cset^{K}_{\{l_i\}}$. As $\cset_{\{l_i\}}$
is a totally ordered set with respect to $\ordering$ which has a global minimum, 
the non-trivial terms of ${\Psi}_{\{l_i\}}$, provided ${\Psi}_{\{l_i\}}$ is non-trivial,
need to have an $\ordering$-minimum $X_0\in\cset^{A}_{\{l_i\}}$. Thus the 
coefficient $k_{X_0}$ of $X_0$ in ${\Psi}_{\{l_i\}}$ must be non-trivial.  As 
$\ordering$ is adapted to $\cset^{A}_{\{l_i\}}$ one can find an annihilation 
operator $\Gamma $ such that $\Gamma X_0\ket{\{\Delta_i\}}$ contains a 
non-trivial term (for a suitably chosen basis depending on $X_0$) that cannot 
be created by $\Gamma $ acting on any other term of ${\Psi}_{\{l_i\}}$ which 
is $\ordering$-larger than $X_0$. But $X_0$ was chosen to be the
$\ordering$-minimum of ${\Psi}_{\{l_i\}}$. Therefore,
$\Gamma X_0\ket{\{\Delta_i\}}$ contains a non-trivial term that cannot be 
created from any other term of ${\Psi}_{\{l_i\}}$. The coefficient of this term 
is obviously given by $ck_{X_0}$ with $c$ a non-trivial complex number. But 
${\Psi}_{\{l_i\}}$ is also a singular vector and must be annihilated by any 
annihilation operator, in particular by $\Gamma$. It follows that $k_{X_0}=0$,
contrary to our original assumption. Thus, the set of non-trivial terms of 
${\Psi}_{\{l_i\}}$ is empty and therefore ${\Psi}_{\{l_i\}}=0$. This results in
$\Psi^{1}_{\{l_i\}} = \Psi^{2}_{\{l_i\}}$.
\eprf

Theorem \refoth{\ref{th:kernel}} states, therefore, that if two singular vectors 
with the same weights, in the same Verma module, agree on the coefficients
of the ordering kernel, then they are identical. If the ordering kernel is trivial 
we consequently find $0$ as the only vector that can satisfy the highest weight 
conditions. This observation provides a proof for Theorem \refoth{\ref{th:striv}}:

\bprf
The trivial vector $0$
satisfies any annihilation conditions for any weights ${\{l_i\}}$. As the 
ordering kernel is trivial the components of the vectors $0$ and $\Psi_{\{l_i\}}$
agree on the ordering kernel and using theorem \refoth{\ref{th:kernel}}
we obtain $\Psi_{\{l_i\}}=0$.
\eprf

Using Theorem \refoth{\ref{th:kernel}} it is also easy to prove Theorem 
\refoth{\ref{th:dims}}:

\bprf
Suppose there were more than $n$ linearly independent singular vectors
$\Psi_{\{l_i\}}$ in $\vm_{\{\Delta_i\}}$ with weights ${\{l_i\}}$. We 
choose $n+1$ linearly independent singular vectors among them 
$\Psi_1$,$\ldots$,$\Psi_{n+1}$. The ordering kernel $\cset^{K}_{\{l_i\}}$ 
has the $n$ elements $X_1$,$\ldots$,$X_n$. Let $k_{jk}$ denote the 
coefficient of the term $X_j$ in the vector $\Psi_k$ in a suitable basis
decomposition. The coefficients $k_{jk}$ thus form a $n$ by $n+1$ matrix 
$M$. The homogeneous system of linear equations $M\lambda=0$ thus has 
a non-trivial solution $\lambda^0=(\lambda^0_1,\ldots,\lambda^0_{n+1})$
for the vector $\lambda$. We then form the linear combination 
$\Psi=\sum_{i=1}^{n+1} \lambda^0_i\Psi_i$. Obviously, the coefficient of $X_j$ 
for the vector $\Psi$  is just given by the $j$-th component of the vector 
$M\lambda$ which is trivial for $j=1,\ldots,n$. Hence, the coefficients of $\Psi$
are trivial on the ordering kernel. On the other hand, $\Psi$ is a linear
combination of singular vectors and therefore it is also a singular vector.
Due to theorem \refoth{\ref{th:kernel}} one immediately finds that 
$\Psi\equiv 0$ and therefore $\sum_{i=1}^{n+1}\lambda_i\Psi_i=0$. This, 
however, contradicts the assumption that $\Psi_1$,$\ldots$, $\Psi_{n+1}$ are 
linearly independent.
\eprf

As a simple example of the Adapted Ordering Method we will see now the  
application of this method to the Virasoro algebra $\vir$, which has been
extensively studied in the literature\cite{Kac,FeFu,RCW}. 
This algebra is 
given by the commutation relations
\bea
\left[L_m,L_n\right] = (m-n) L_{m+n} +\frac{C}{12} (m^3-m) \delta_{m+n,0}
\,, & \left[C,L_m\right] = 0 \,, & m,n\in\bbbz \,,
\eea
where $C$ commutes with all operators of $\vir$ and can hence be taken to be 
constant $c\in\bbbc$. $\vir$ can be written in its {\it triangular decomposition}:
$\vir=\vir^-\oplus \vir^0 \oplus \vir^+$, where $\vir^+=\spn{L_m:m\in{\bf {N}}}$ 
is the set of {\it annihilation operators}, $\vir^-=\spn{L_{-m}:m\in\bbbn}$ is the 
set of {\it creation operators}, and the {\it Cartan subalgebra} is given by 
$\vir^0=\spn{L_0,C}$. For elements of $\vir$ that are eigenvectors of $L_0$ 
with respect to the adjoint representation the $L_0$-eigenvalue is called the 
{\it level} $l$. For the universal enveloping algebra $U(\vir)$, elements of the 
form $L_{-p_I}\ldots L_{-p_1}$, $p_q\in\bbbz$ for $q=1,\ldots, I$, 
$I\in\bbbn$, are at level $l=\sum_{q=1}^{I}p_q$. Note that annihilation 
operators $L_m \in \vir^+$ have negative level $l= -m$.

A representation with $L_0$-eigenvalues bounded from below contains a 
vector with $L_0$-eigenvalue $\Delta$ which is annihilated by $\vir^+$, a
{\it highest weight (h.w.) vector} $\ket{\Delta}$:
\bea
\vir^+ \ket{\Delta}=0\,, & L_0\ket{\Delta}=\Delta\ket{\Delta}\,. \label{eq:hwc}
\eea
The {\it Verma module} $\vm_{\Delta}$ built on $\ket{\Delta}$
is $L_0$-graded in a natural way. The corresponding $L_0$-eigenvalue 
is called the {\it conformal weight} and is written for convenience as $\Delta+l$, 
where $l$ is the {\it level}. Any proper submodule of $\vm_{\Delta}$ needs 
to contain a {\it singular vector} $\Psi_l$ that is not proportional to the 
h.w. vector $\ket{\Delta}$ but still satisfies the {\it h.w. vector 
conditions} with conformal weight $\Delta+l$: 
\bea
\vir^+\Psi_l=0 \,, & L_0\Psi_l=(\Delta+l)\Psi_l \,. \label{eq:hwc2}
\eea

Now we will see the total ordering on $\cset_l$ defined by Kent\cite{adrian1} 
for the Virasoro algebra. One has to take into account, however, that Kent used 
the following ordering to show that in his generalised Verma modules vectors at 
level $0$ satisfying the h.w. conditions are actually proportional to the 
h.w. vector. Using the Adapted Ordering technology, though, one 
deduces that this ordering already implies that all Virasoro singular vectors are 
unique at their levels up to proportionality, simply because the ordering kernel 
for each $l\in\bbbn$ has just one element: $L_{-1}^l$.

\bdf \label{def:virorder}
On the set $\cset_l$ of Virasoro operators we introduce the total ordering 
$\ordering_{\vir}$ for $l\in\bbbn$. For two elements $X_1,X_2\in\cset_l$, 
$X_1\neq X_2$, with $X_i=L_{-m^i_{I_i}}\ldots L_{-m^i_1}L_{-1}^{n^i}$,
$n^i=l-m^i_{I_i}\ldots-m^i_1$, or $X_i=L_{-1}^l$, $i=1,2$ we define
\bea
X_1 \osm{\ordering_{\vir}}X_2 & {\rm if} & n^1>n^2\,. \label{eq:virord1}
\eea
If, however, $n^1=n^2$ we compute the index
$j_0=\min\{j:m^1_j-m^2_j\neq 0,j=1,\ldots,\min(I_1,I_2)\}$. We then define
\bea
X_1 \osm{\ordering_{\vir}}X_2 & {\rm if} & m^1_{j_0}<m^2_{j_0}\,.
\eea
For $X_1=X_2$ we set $\, X_1 \osm{\ordering_{\vir}}X_2$ and
$\, X_2\osm{\ordering_{\vir}}X_1$.
\edf
The index $j_0$ describes the first mode, read from the right to the left, for 
which the generators in $X_1$ and $X_2$ ($L_{-1}$ excluded) are different. 
For example, in $\cset_8$ one has  $L_{-2}L_{-2}L_{-2}L_{-1}^2 
\osm{\ordering_{\vir}} L_{-4}L_{-2}L_{-1}^2$ with index $j_0=2$. Observe that 
$L_{-1}^l\in\cset_l$ is the global $\ordering_{\vir}$-minimum in $\cset_l$. Now 
using the Adapted Ordering Method one finds the following theorem\cite{SD1}. 

\bth \label{th:viradap}
The ordering $\ordering_{\vir}$ is adapted to $\cset^A_l=
\cset_l\setminus\{L_{-1}^l\}$ for each level $l\in\bbbn$ and
for all Verma modules $\vm_{\Delta}$. The ordering kernel is
given by the single element set $\cset^K_l=\{L_{-1}^l\}$.
\eth
For example let us consider the set of terms at level 3, 
$\cset_3=\{L^3_{-1}, L_{-2}L_{-1}, L_{-3}\}$. 
One finds the total ordering 
$L^3_{-1} \osm{\ordering_{\vir}} L_{-2}L_{-1} \osm{\ordering_{\vir}} L_{-3}$,
which is adapted to $\cset^A_3=\{L_{-2}L_{-1}, L_{-3}\}$ with the ordering
kernel $\cset^K_3=\{L_{-1}^3\}$. To see this one has to compute 
the action of the annihilation operators $\Gamma \in \{L_1, L_2, L_3\}$ 
on the three terms. In fact, the action of $L_1$ already reveals the
structure of $\cset^A_3$, as $L_1 L_{-2} L_{-1} \ket{\Delta}$ contains 
the term $L^2_{-1}$ that is absent in $L_1 L_{-3} \ket{\Delta}$. The
action of the three annihilation operators on $L^3_{-1} \ket{\Delta}$,
however, produce terms that are also created by the action of these
operators on $L_{-2} L_{-1} \ket{\Delta}$ and/or $L_{-3} \ket{\Delta}$.

Finally, from the previous theorem one now deduces the known result about
the uniqueness of the Virasoro singular vectors\cite{adrian1}.

\bth \label{th:viruni}
If the Virasoro Verma module $\vm_{\Delta}$ contains a singular vector 
$\Psi_l$ at level $l$, $l\in\bbbn$, then $\Psi_l$ is unique up to proportionality. 
\eth


\section{Results for the superconformal algebras}
\label{sec:Nis2}

As an illustration of the power of the Adapted Ordering Method, in this section we will 
review briefly the results obtained for the $N=2$ and Ramond $N=1$ superconformal 
algebras. This method has been applied to the topological, to the Neveu-Schwarz
and to the Ramond  $N=2$ algebras in Ref. \icite{SD1}, to the twisted $N=2$ algebra 
in Ref. \icite{SD2} and to the Ramond $N=1$ algebra in Ref. \icite{Ramond}. 
As the representation theory of these superconformal algebras
has different types of Verma modules, one has to introduce different adapted 
orderings for each type and the corresponding kernels also allow different 
degrees of freedom. 

Let us start with the topological $N=2$ superconformal algebra $\ta$. 
It contains the Virasoro generators $\cL_m$ with trivial central extension,
a Heisenberg algebra $\cH_m$ corresponding to the U(1) current, and the
fermionic generators $\cG_m$ and $\cQ_m$ corresponding to
two anticommuting fields with conformal weights 2 and 1 respectively.
$\ta$ satisifies the (anti-)commutation relations\cite{DVV}

\be
\begin{array}{ll}
\left[\cL_m,\cL_n\right] = (m-n)\cL_{m+n}\,, &
\left[\cH_m,\cH_n\right] = \frac{C}{3}m\delta_{m+n}\,,\\
\left[\cL_m,\cG_n\right] = (m-n)\cG_{m+n}\,, &
\left[\cH_m,\cG_n\right] = \cG_{m+n}\,,\\
\left[\cL_m,\cQ_n\right] = -n\cQ_{m+n}\,, &
\left[\cH_m,\cQ_n\right] = -\cQ_{m+n}\,,\\
\left[\cL_m,\cH_n\right] = -n\cH_{m+n}+\frac{C}{6}(m^2+m)\delta_{m+n}\,,\\
\left\{\cG_m,\cQ_n\right\} = 2\cL_{m+n}-2n\cH_{m+n}
+\frac{C}{3}(m^2+m)\delta_{m+n}\,,\\
\left\{\cG_m,\cG_n\right\} = \left\{\cQ_m,\cQ_n\right\} = 0\,, &
 m,~n\in\bbbz\,.\label{topalgebra}
\end{array}
\ee 
The set of {\it annihilation operators} $\ta^+$ is spanned by the generators 
with positive index, the set of {\it creation operators} $\ta^-$ is spanned by 
the generators with negative index, and the {\it zero modes}  are given by 
$\ta^0=\spn{\cL_0,\cH_0,C,\cG_0,\cQ_0}$. The Cartan subalgebra 
is generated by ${\cal H}_{\ta}=\spn{\cL_0,\cH_0, C}$, where $C$ can 
be taken to be constant $c\in\bbbc$, and the fermionic generators 
$\{\cG_0,\cQ_0\}$ classify the different choices of Verma modules.

A h.w. vector $\ket{\Delta,h}^{\cal N}$ is an eigenvector of ${\cal H}_{\ta}$ 
with $\cL_0$ eigenvalue $\Delta$, $\cH_0$ eigenvalue $h$, and vanishing
$\ta^{+}$ action. Additional vanishing conditions ${\cal N}$ are possible with 
respect to the operators $\cG_0$ and $\cQ_0$, resulting as 
follows\cite{beatriz2}. One can distinguish four different types of h.w.
vectors $\ket{\Delta,h}^{\cal N}$ labeled by a superscript
${\cal N}\in\{G,Q,GQ\}$, or no superscript at all: h.w. vectors 
$\ket{\Delta,h}^{G}$ annihilated by $\cG_0$ but not by $\cQ_0$  
($\cG_0$-closed), h.w. vectors $\ket{\Delta,h}^{Q}$ annihilated 
by $\cQ_0$ but not by $\cG_0$ ($\cQ_0$-closed), h.w. vectors
$\ket{0,h}^{GQ}$ annihilated by both $\cG_0$ and $\cQ_0$ ({\it chiral}), 
with zero conformal weight necessarily, and finally undecomposable 
h.w. vectors  $\ket{0,h}$ that are neither annihilated by 
$\cG_0$ nor by $\cQ_0$  ({\it no-label}), also with zero conformal weight.
Hence we have four different types of Verma modules\cite{beatriz2}:
$\vm_{\Delta,h}^G$, $\vm_{\Delta,h}^Q$, $\vm_{0,h}^{GQ}$ and 
$\vm_{0,h}$, built on the four different types of h.w. vectors. 

For elements $X$ of $\ta$ which are eigenvectors of ${\cal H}_{\ta}$ with 
respect to the adjoint representation one defines the {\it level} $l$ as 
$[\cL_0,X]= l X$ and the {\it charge} $q$ as $[\cH_0,X]= q X$. In particular, 
elements of the form
$X=\cL_{-l_L} \ldots \cL_{-l_1}\cH_{-h_H} \ldots \cH_{-h_1}
\cQ_{-q_Q} \ldots \cQ_{-q_1}\cG_{-g_G} \ldots \cG_{-g_1}$,
and any reorderings of it, have level $l=\sum_{j=1}^{L}l_j
+\sum_{j=1}^{H}h_j+\sum_{j=1}^{Q}q_j+\sum_{j=1}^{G}g_j$ and
charge $q=G-Q$. The Verma modules are naturally
$\bbbn_0\times\bbbz$ graded with respect to the ${\cal H}_{\ta}$
eigenvalues relative to the eigenvalues $(\Delta,h)$ of the h.w. vector. 
For a vector $\Psi_{l,q}$ in $\vm_{\Delta,h}^N$ the $\cL_0$-eigenvalue 
is $\Delta+l$ and the $\cH_0$-eigenvalue is $h+q$ with the level 
$l\in\bbbn_0$ and the relative charge $q\in\bbbz$. 

The singular vectors are annihilated by $\ta^+$ and may also satisfy additional 
vanishing conditions with respect to the operators $\cG_0$ and $\cQ_0$. 
Therefore one also distinguishes singular vectors of the types\cite{beatriz2} 
$\Psi^G_{l,q}$, $\Psi^Q_{l,q}$, $\Psi^{GQ}_{l,q}$ and $\Psi_{l,q}$. 
As there are $4$ types of Verma modules and $4$ types of singular vectors 
one might think of $16$ different combinations of singular vectors in Verma 
modules. However, no-label and chiral singular vectors do not exist neither in 
{\it chiral} Verma modules $\vm_{0,h}^{GQ}$ nor in {\it no-label} Verma 
modules\cite{beatriz2} $\vm_{0,h}$ (with one exception: 
chiral singular vectors exist at level 0 in no-label Verma modules).
Using the Adapted Ordering Method one has to introduce adapted orderings
for the remaining $12$ combinations, whose kernels give upper limits for
the dimensions of the corresponding singular vector spaces. One finds that 
for most charges $q$ singular vectors do not exist. For the case of the
Verma modules $\vm_{\Delta,h}^G$ built on $\cG_0$-closed
h.w. vectors $\ket{\Delta,h}^G$, for $c\neq 3$, the maximal
dimensions for the singular vector spaces $\Psi_{l,q}^{G}$, $\Psi_{l,q}^{Q}$
$\Psi_{l,q}^{GQ}$ and $\Psi_{l,q}$ are given as follows\cite{SD1}:

\btab{|l|c|c|c|c|c|}
\hline \label{tab:dim1}
 & $q=-2$ & $q=-1$ & $q=0$ & $q=1$ & $q=2$ \\
\hline
$\Psi_{l,q,\ket{\Delta,h}^G}^{G}$ &
$0$ & $1$ & $2$ & $1$ & $0$ \\
\hline
$\Psi_{l,q,\ket{\Delta,h}^G}^{Q}$ &
$1$ & $2$ & $1$ & $0$ & $0$ \\
\hline
$\Psi_{l,q,\ket{-l,h}^G}^{GQ}$ &
$0$ & $1$ & $1$ & $0$ & $0$ \\
\hline
$\Psi_{l,q,\ket{-l,h}^G}$ &
$0$ & $1$ & $1$ & $0$ & $0$ \\
\hline
\etab{Maximal dimensions for singular vectors spaces in $\vm_{\Delta,h}^G$.}
Charges $q$ that are not given have dimension $0$ and hence do not allow 
any singular vectors. The maximal dimensions for the case of the Verma 
modules $\vm_{\Delta,h}^Q$, for $c\neq 3$, are obtained simply by 
interchanging $G \leftrightarrow Q$ and $q \leftrightarrow -q$ in the previous 
table.

For the case of singular vectors in chiral Verma modules $\vm_{0,h}^{GQ}$ 
and in no-label Verma modules $\vm_{0,h}$, for $c\neq 3$, one obtains 
the following maximal dimensions\cite{SD1}:

\btab{|l|c|c|c|c|c|}
\hline \label{tab:dim3}
 & $q=-2$ & $q=-1$ & $q=0$ & $q=1$ & $q=2$ \\
\hline
$\Psi_{l,q,\ket{0,h}^{GQ}}^{G}$ &
$0$ & $0$ & $1$ & $1$ & $0$ \\
\hline
$\Psi_{l,q,\ket{0,h}^{GQ}}^{Q}$ &
$0$ & $1$ & $1$ & $0$ & $0$ \\
\hline
$\Psi_{l,q,\ket{0,h}}^{G}$ &
$0$ & $1$ & $3$ & $3$ & $1$ \\
\hline
$\Psi_{l,q,\ket{0,h}}^{Q}$ &
$1$ & $3$ & $3$ & $1$ & $0$ \\
\hline
\etab{Maximal dimensions for singular vectors spaces
in $\vm_{0,h}^{GQ}$ and in $\vm_{0,h}$.}

Tables \tab{\ref{tab:dim1}} and \tab{\ref{tab:dim3}} prove the conjecture 
made in Ref. \icite{beatriz2}, using the algebraic mechanism denoted 
{\it the cascade effect}, about the possible existing types of topological
singular vectors.  In addition, low level examples were 
constructed\cite{beatriz2} for all these types, what proves that all 
of them exist already at level 1. The four types of two-dimensional 
singular vector spaces of \tab{\ref{tab:dim1}} also exist starting at level 2, 
and four examples at level 3 were constructed\cite{beatriz2} as well.
For the case of the three-dimensional singular vector spaces in no-label 
Verma modules in \tab{\ref{tab:dim3}}, the corresponding types of singular 
vectors have been constructed at level 1 generating 
one-dimensional\cite{beatriz2} as well as two-dimensional\cite{SD1} spaces,
but no further search has been done for the three-dimensional spaces.

Transferring the dimensions we have found in tables \tab{\ref{tab:dim1}} 
and \tab{\ref{tab:dim3}} to the Neveu-Schwarz $N=2$ 
algebra\cite{ademollo,BFK,Dobrev2,Matsuo,Kiritsis} is
straightforward as this algebra is related to the topological $N=2$ 
algebra through the topological twists ${\ }T_W^{\pm}$:
${\ }\cL_m=L_m\pm 1/2 H_m$, ${\ }\cH_m=\pm H_m$,
${\ }\cG_m=G^{\pm}_{m+1/2}$ and ${\ }\cQ_m=G^{\mp}_{m-1/2}$, 
where $G^{\pm}_{m+1/2}$ are the half-integer moded fermionic 
generators. As a result, the standard Neveu-Schwarz h.w. vectors
correspond to $\cG_0$-closed topological h.w. vectors, whereas the 
chiral (antichiral) Neveu-Schwarz h.w. vectors, annihilated by $G^+_{-1/2}$ 
($G^-_{-1/2}$), correspond to chiral topological h.w. vectors.
This implies\cite{beatriz1,beatriz2} that the standard, chiral and antichiral 
Neveu-Schwarz singular vectors correspond to topological singular vectors 
of the types $\Psi^{G}_{l,q,\ket{\Delta,h}^G}$ 
and $\Psi^{GQ}_{l,q,\ket{\Delta,h}^G}$, whereas the ones built in chiral 
or antichiral Verma modules correspond to topological singular 
vectors of only the type $\Psi^{G}_{l,q,\ket{\Delta,h}^{GQ}}$.
As a consequence, by untwisting the first row of table \tab{\ref{tab:dim1}} 
one recovers the results\cite{thesis,paper2} that in Verma modules of the 
Neveu-Schwarz $N=2$ algebra singular vectors only exist for charges 
$q=0, \pm 1$ and two-dimensional spaces only exist for uncharged singular
vectors. By untwisting the third row of table \tab{\ref{tab:dim1}} one gets
a proof for the conjecture\cite{beatriz2} that chiral singular vectors in 
Neveu-Schwarz Verma modules only exist for $q=0, 1$ whereas antichiral 
singular vectors only exist for $q=0, -1$. The untwisting of the first row
of table \tab{\ref{tab:dim3}}, finally, proves the 
conjecture\cite{beatriz1,beatriz2}
that in chiral Neveu-Schwarz Verma modules $\,\vm_{h/2,h}^{NS,ch}\,$ 
singular vectors only exist for $q=0, -1$, whereas in antichiral Verma 
modules $\,\vm_{-h/2,h}^{NS,a}\,$ singular vectors only exist for $q=0, 1$.

As to the representations of the Ramond $N=2$ 
algebra\cite{BFK,Dobrev2,Matsuo,Kiritsis}, they 
are exactly isomorphic to the representations of the topological $N=2$ 
algebra. Namely, combining the topological twists ${\ }T_W^{\pm}$ and the 
spectral flows one constructs a one-to-one mapping between
the Ramond singular vectors and the topological singular vectors, at
the same levels and with the same charges\cite{DGR3}. Therefore the
results of tables \tab{\ref{tab:dim1}}  and \tab{\ref{tab:dim3}} can be
transferred to the Ramond singular vectors simply by exchanging the labels
$G \to (+), {\ }Q \to (-)$, where the helicity $(\pm)$ denotes the vectors 
annihilated by the fermionic zero modes $G_0^{\pm}$, and by taking
into account that the chiral and undecomposable {\it no-helicity} Ramond
vectors\cite{beatriz2,DGR1,DGR3}, require conformal weight 
$\Delta+l=c/24 \,$. 

The twisted $N=2$ superconformal algebra\cite{BFK,Dobrev2,Matsuo,Kiritsis}
is not related to the other three $N=2$ algebras. It has mixed modes, integer 
and half-integer, for the fermionic generators, and the eigenvectors have
no charge, as the U(1) current generators are half-moded, but they have
{\it fermionic parity}. The Adapted Ordering Method was worked out for
the twisted $N=2$ algebra in Ref. \icite{SD2}. The maximal dimension for 
the singular vector spaces in standard Verma modules was found to be two 
and these two-dimensional singular spaces were shown to exist by explicit 
computation, starting at level $3/2$. In Verma modules built on $G_0$-closed
h.w. vectors, however, the singular vectors were found to be only
one-dimensional. This method also allowed to propose a reliable conjecture
for the coefficients of the relevant terms of all singular vectors, i.e. for
the coefficients with respect to the ordering kernels, what made possible
to identify all the cases of two-dimensional singular vector spaces for all
levels, as well as to identify all $G_0$-closed singular vectors.
The resulting expressions, in turn, led to the discovery of 
subsingular vectors for this algebra, and several explicit examples were
also computed. Finally, the multiplication rules for singular vectors
operators were derived using the ordering kernel coefficients, what set
the basis for the analysis of the twisted $N=2$ embedding diagrams.    

Finally let us consider the $N=1$ superconformal
algebras\cite{Kac,Kiritsis,MRC,Dobrev1}. 
The structure of the h.w. representations of the Neveu-Schwarz $N=1$ 
algebra has been completely understood in Ref. \icite{Astash}. The 
corresponding Verma modules do not contain two-dimensional singular vector 
spaces neither subsingular vectors. In the case of the Ramond $N=1$ algebra,
however, the application of the Adapted Ordering Method in Ref. \icite{Ramond}
has shown that its representations have a much richer structure than previously
suggested in the literature. In particular, it was found that standard Verma modules
may contain two-dimensional singular vector spaces and also subsingular vectors.
Moreover, the two-dimensional ordering kernels allowed to derive multiplication
rules for singular vector operators and led to expressions for two-dimensional
singular spaces. Using these multiplication rules descendant singular vectors were
studied and embedding diagrams were derived for the rational models. In addition,
this allowed to conjecture the ordering kernel coefficients of all singular vectors
and therefore identify these vectors uniquely.   

\section{Conclusions and Final Remarks}

We have presented the Adapted Ordering Method for general algebras and 
superalgebras provided they can be triangulated, like is the case for most
Lie algebras and superalgebras.  This method is based on the concept of
adapted ordering, which implies that any singular vector needs to contain at 
least one non-trivial term included in the ordering kernel. The size of the 
ordering kernel therefore limits the dimension of the corresponding singular
vector space. As a result the adapted orderings must be chosen such that 
the ordering kernels are as small as possible. Weights for which the ordering
kernels are trivial do not allow any singular vectors in the corresponding weight
spaces. On the other hand, non-trivial ordering kernels give us the maximal
dimension of a possible singular vector space and uniquely define all singular
vectors through the coefficients with respect to them. 

The Adapted Ordering 
Method has been applied so far to the $N=2$ and Ramond $N=1$ 
superconformal algebras, allowing to prove several conjectured results as
well as to obtain many new results, as we have reviewed. For example,
this method allowed to discover subsingular vectors and two-dimensional 
spaces of singular vectors for the twisted $N=2$ and Ramond $N=1$ 
algebras\cite{SD2,Ramond}. (For the other three isomorphic $N=2$ algebras
two-dimensional singular spaces had been discovered\cite{beatriz2,thesis,paper2},
as well as subsingular vectors\cite{subsing,beatriz1,DGR1,beatriz2}, before the
Adapted Ordering Method was applied to them). We are convinced therefore that
this method should be of very much help for the study of the representation
theory of many other algebras, in particular the $N>2$ superconformal algebras.

\vskip .5in
\acknowledgements

I am grateful to Christoph Schweigert for providing some information about
triangulated algebras and to Bert Schellekens for reading carefully the
manuscript.
The work of the author is partially supported by funding of the spanish
Ministerio de Educaci\'on y Ciencia, Research Project BFM2002-03610.

\noi



\end{document}